\documentclass[multphys,vecphys]{svmult}

\usepackage{makeidx}         
\usepackage{graphicx}        
\usepackage{multicol}        
\usepackage[bottom]{footmisc}
\usepackage{amssymb}

\begin{document}

\title*{A Search for \textit{Deep Impact}'s Large Particle Ejecta}
\author{Michael S. Kelley,\inst{1} William T. Reach\inst{2}, \and
Charles E. Woodward\inst{1}}
\institute{Department of Astronomy, University of Minnesota, 116
Church St SE, Minneapolis, MN, 55455 \texttt{msk@astro.umn.edu} \and
Spitzer Science Center, MS 220-6, California Institute of Technology,
Pasadena, CA 91125}
%
%
\maketitle

%
%

\setcounter{footnote}{0}

%
%

\begin{abstract}
The \textit{Deep Impact} encounter with the nucleus of 9P/Tempel
ejected small grains ($a \lesssim 10$~\mbox{$\mu$m}) into the comet's coma,
evidenced by thermal emission from small dust grains at mid-infrared
wavelengths ($\lambda \sim 10$~\mbox{$\mu$m}{}) and dynamical simulations of
optical images.  Meteor-sized particles ($a \gtrsim 100$~\mbox{$\mu$m})
ejected by the impact will likely have the lowest ejection velocities
and will weakly interact with solar radiation pressure.  Therefore,
large particles may remain near the nucleus for weeks or months after
ejection by \textit{Deep Impact}.  We present initial highlights of
our \textit{Spitzer Space Telescope}/MIPS 24~\mbox{$\mu$m}{} camera program
to image comet 9P/Tempel at 30, 80, 420, and 560 days after the
\textit{Deep Impact} encounter.  The MIPS data, combined with our
dynamics model, enable detection of large dust grains potentially
ejected by \textit{Deep Impact}.
\end{abstract}
\section{Introduction}
The \textit{Deep Impact} impactor released an abundance of sub-micron
sized particles into the coma of 9P/Tempel.  Mid-infrared spectra of
the ejecta displayed a strong silicate emission feature, including
crystalline silicates, indicating the presence of small particles
\cite{harker05, sugita05}.  Ground-based images of the ejecta showed
dust that strongly interacted with radiation pressure, again implying
small grains \cite{schleicher06b}.  This small particle ejecta plume
was optically thick, apparent from a shadow cast by the ejecta onto
the comet nucleus \cite{ahearn05}.  Though the impact site was
available for observation by the \textit{Deep Impact} flyby spacecraft
for 800~s after impact, the crater was wholly obscured by the
optically thick ejecta.  Had the crater been observed, a size estimate
of the crater, and therefore, an estimate of the total ejecta mass,
may have been directly determined.  Instead, we must estimate the
total ejecta mass from observations of the ejecta itself.  The gas and
small particle dust (0.1--10~\mbox{$\mu$m}{}) masses have been constrained
to be of order $10^5$--$10^6$~kg \cite{kueppers05, keller05, harker05,
sugita05}, but the mass in larger sized dust grains ($\gtrsim
100$~\mbox{$\mu$m}) is unknown.  If the grains were ejected with typical
comet grain size distributions ($dn/da \sim a^{-3.5}$), the large dust
mass could dominate the total ejected dust mass.

The particle size distribution may have strongly favored small
particles; therefore, the surface brightness of large grains would be
faint and difficult to detect in contrast with the smaller, more
abundant, grains.  Schleicher et al.{} \cite{schleicher06b}, through
dynamic modeling of ground-based optical images, estimated that the
effective radii of ejected particles ranged from 0.25~to
1.25~\mbox{$\mu$m}{} with ejection velocities of 0.13--0.23~km~s$^{-1}$.  Dynamical
simulations of larger particles did not agree with the observations,
yet this does not preclude their existence since they may be many
times less abundant than the small particles.  Moreover, there is some
evidence for small and large particle segregation in the ejecta
\cite{sugita06}, likely due to the ejection of larger particles with
lower ejection velocities.  If the ejected grain size distribution
extends beyond 10--100~\mbox{$\mu$m}, then for $v_{ej}<0.1$~km~s$^{-1}${} the
largest particles could remain near the nucleus for days, or even
weeks after impact.  High sensitivity, long temporal baseline
observations, beyond the immediate aftermath of \textit{Deep Impact},
are necessary to observe any large dust grains generated by crater
formation processes.  The slowest particles likely remained
gravitationally bound to the nucleus and fell back to the surface,
therefore, such observations will only provide a lower limit to the
total excavated mass.  We present dynamical simulations and
\textit{Spitzer}/MIPS images to search for large dust grains in the
\textit{Deep Impact} ejecta.

\section{Observations and Models}

\begin{figure}
\centering
\includegraphics[height=0.6\textwidth,angle=90]{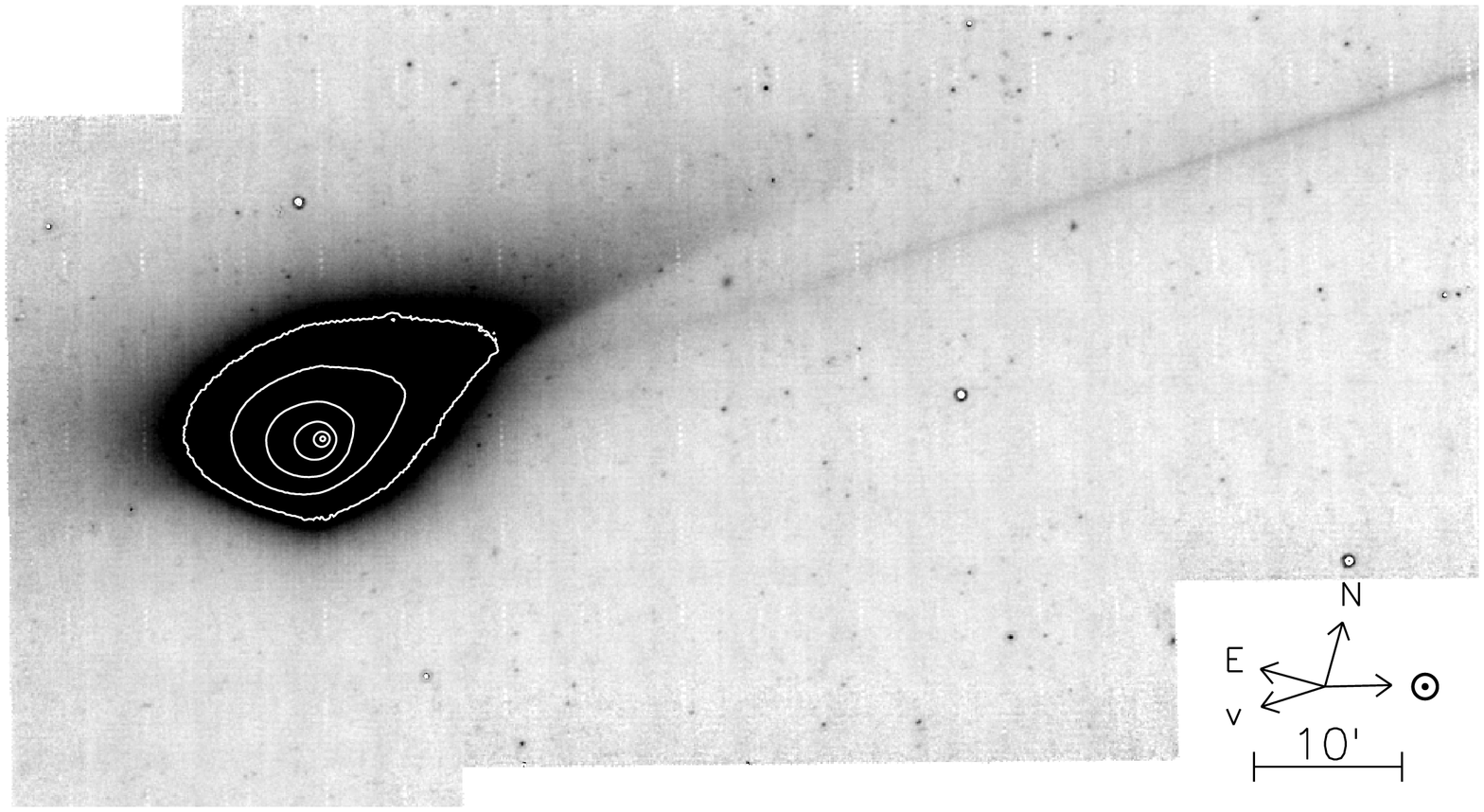}
\includegraphics[height=0.6\textwidth,angle=90]{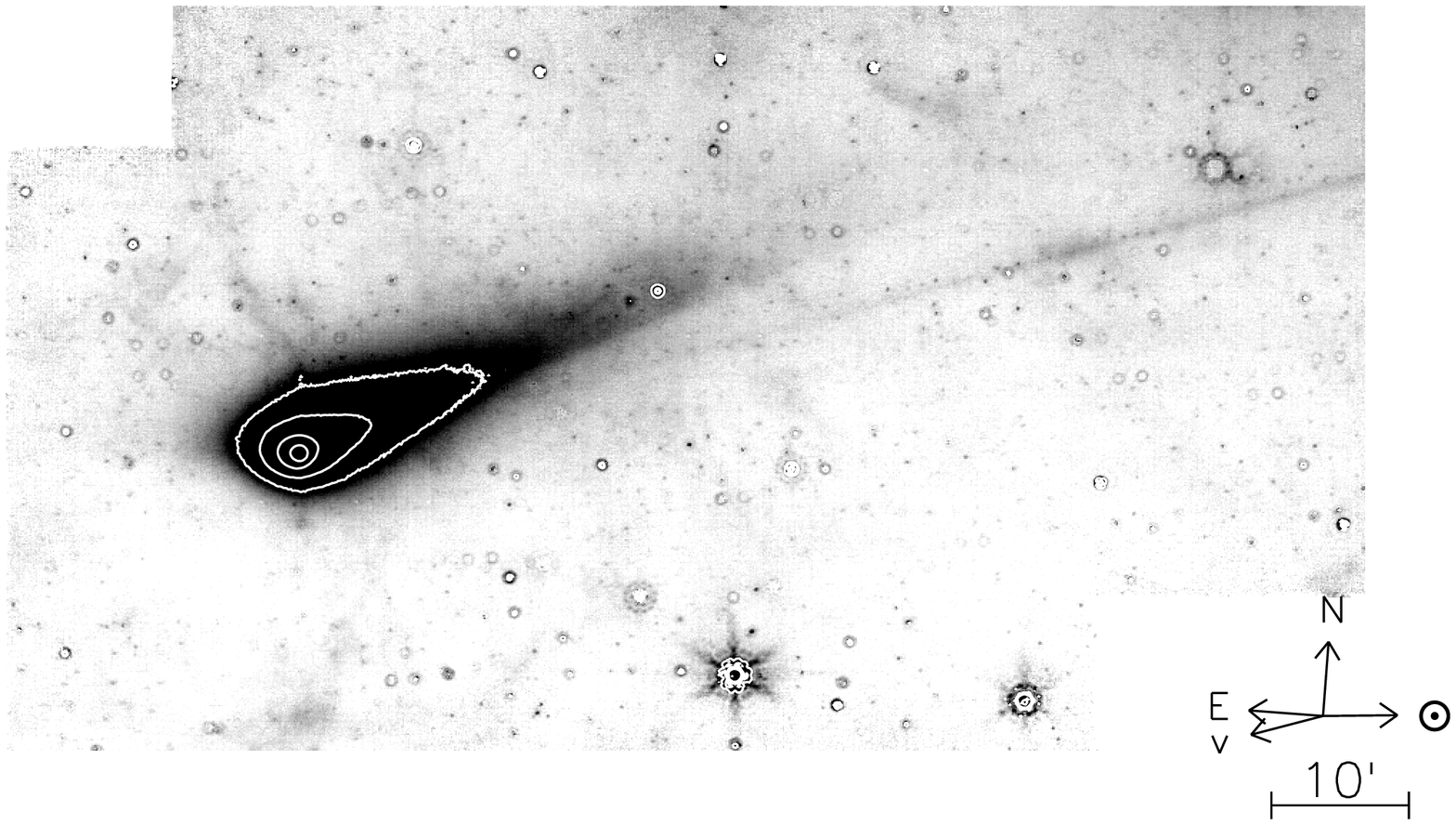}
\caption{\textit{Spitzer}/MIPS 24~\mbox{$\mu$m}{} images of comet 9P/Tempel.
The lowest contour is placed at 3~MJy~sr$^{-1}$, and the contours are
spaced at factor of 3 intervals.  Celestial N and E, the projected
sunward direction ($\odot$), and comet 9P's projected velocity vector
(v) are provided for each image.  The comet's natural dust trail is
seen extending from the comet in the negative velocity direction.}
\label{kelley:fig:1}
\end{figure}

We initiated an observing program with the \textit{Spitzer Space
Telescope} \cite{werner04} to observe comet 9P/Tempel with the MIPS
24~$\mu$m camera \cite{rieke04} at 30, 80, 420, and 560~days after the
\textit{Deep Impact} encounter.  Two observations, obtained on
2005~Jul~31 and on 2005~Sep~23, have been processed and analyzed (see
Table~\ref{kelley:table:obs} and Fig.~\ref{kelley:fig:1}).  The images
span a $22^\prime\times42^\prime$ area surrounding the nucleus and
dust trail.  The dust trail is visible in both images as the thin,
isolated, linear feature extending from the inner coma to the
northwestern edges.

\begin{table}
\begin{center}
\caption{\textit{Spitzer}/MIPS 24~\mbox{$\mu$m}{} post-\textit{Deep Impact}
observations of comet 9P/Tempel.}
\label{kelley:table:obs}       
\begin{tabular}{cccccc}
\hline\noalign{\smallskip}
& Time & $\Delta T$ & $r_h$ & $\Delta_s$ & Phase Angle \\
Date & (UT) & (days)$^{\mbox{a}}$ & (AU) & (AU)$^{\mbox{a}}$ & (degrees) \\
\noalign{\smallskip}\hline\noalign{\smallskip}
2005~Jul~31 & 04:14 & 26.96 & 1.53 & 0.87 & 40 \\
2005~Sep~23 & 21:00 & 81.66 & 1.68 & 1.31 & 38 \\
2006~Sep~01 & 08:25 & 424.11 & 3.52 & 3.15 & 17 \\
2007~Jan~15$^{\mbox{c}}$ & 00:00 & 560 & 4.03 & 3.55 & 13 \\
\noalign{\smallskip}\hline
\end{tabular}
\end{center}

$^{\mbox{a}}$Time elapsed since impact.

$^{\mbox{b}}$Comet-\textit{Spitzer} distance.

$^{\mbox{c}}$Estimated date of observation.
\end{table}

Images of comet 9P/Tempel following the collision with \textit{Deep
Impact} showed a short lived plume of high $\beta$ particles that
quickly dispersed from the vicinity of the comet \cite{meech05,
keller05, schleicher06b}.  Schleicher et al.{} \cite{schleicher06b}
suggest (based on images 6~days after impact) that very slow moving
particles, $v_{ej}<0.001$~km~s$^{-1}$, may exist.  To predict the location of
the large, slow moving, \textit{Deep Impact}-ejected dust grains in
the MIPS images, we will work under the assumption that the high
$\beta$ ejecta plume has a low $\beta$ component, which was ejected in
the same manner, but with lower velocities.  The simulations will be
created with our Monte-Carlo dynamical model \cite{kelley06}.

Schleicher et al.{} \cite{schleicher06b} reproduced the ejecta plume
using particles with $\beta \approx0.24$--1.2 ($a \approx
0.25$--1.25~\mbox{$\mu$m}{}) and ejection velocities of 0.13 to 0.23~km~s$^{-1}$.
Their best model ejected particles in a cone with an opening angle of
70$^\circ$ and a wall thickness of 30$^\circ$.  The cone was centered
at a position angle of 255$^\circ$ and a ``phase angle'' of 70$^\circ$
(0$^\circ$ is directed toward the Earth, 180$^\circ$ is directly
away).  We begin modeling Schleicher et al.'s ejection plume with
these parameters and verify the parameters by comparison to their
images; however, with our model, the ejecta position angle of
225$^\circ$ reported by Sugita et al.{} \cite{sugita05} produced
better agreement with the observations.  The simulated images (not
shown) exhibit a limb brightened cone at $t_{impact} + 0.97$~days.
The limb brightening persists through each of the time steps.  Limb
brightening is not evident in the images of Schleicher et al.{}
\cite{schleicher06b}.  Although optical images from the \textit{Hubble
Space Telescope} ($t_{impact} + 1$--2~hrs) \cite{feldman06}, mid-IR
images from the Subaru Telescope ($t_{impact} + 3$~hrs)
\cite{sugita05}, and optical images by the \textit{Rosetta} spacecraft
($t_{impact} + 21$~hrs) \cite{keller05} show what may be limb
brightening in the ejecta plume.  Further modeling and interpretation
will be required to resolve this discrepancy.

The evidence for velocity differentiation in the ejected dust is
apparent in the optical images of Schleicher et al.{}
\cite{schleicher06b}, the mid-infrared images of Sugita et al.{}
\cite{sugita05}, and the mid-infrared spectra of Harker et al.{}
\cite{harker06, harker07}.  We posit that there exits a slow moving,
large particle component to the ejecta.  We modeled the slow moving
component with the same cone parameters as the high $\beta$ grains,
but with lower $\beta$-values ($\beta \leq 0.1$) and lower ejection
velocities ($v_{ej} \leq 0.1$~km~s$^{-1}$).  The particles are ejected
at the time of impact with a uniform distribution of velocities
(between 0 and 0.1~km~s$^{-1}$) and a logarithmic distribution of
grain $\beta$ values ($dn/d\log{\beta} \propto 1$) ranging from
$10^{-4}$ to $10^{-1}$ ($a \approx 6-6000$~\mbox{$\mu$m}).  The
logarithmic distribution of $\beta$-values ensures all grain size
decade bins are equally represented (i.e., there is an equal number of
10~\mbox{$\mu$m}{} grains as there are 1000~\mbox{$\mu$m}{} grains).
The model simulated the ejection of $10^6$ particles in a 180$^\circ$
cone centered on the derived impact site, as discussed above.  Those
particles outside of the Schleicher et al.{} \cite{schleicher06b} cone
were removed from the simulated images.

\section{Discussion}
\begin{figure}
\centering
\includegraphics[width=0.8\textwidth]{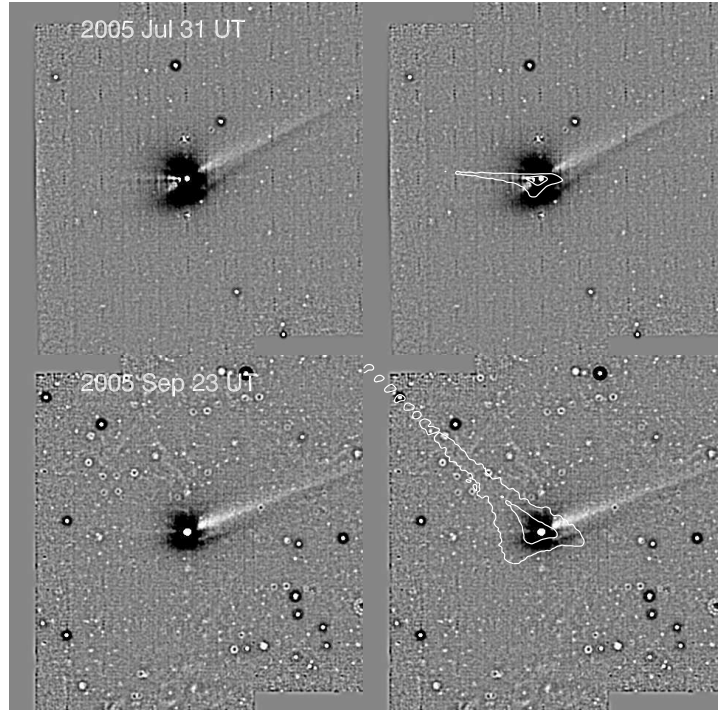}
\caption{(\textit{Left}) \textit{Spitzer}/MIPS images of comet
9P/Tempel enhanced with an unsharp mask to show asymmetries in the
comet coma (20$^\prime$ field-of-view).  (\textit{Right}) Contours
from our large particle simulations are overlayed on the enhanced MIPS
images to show the possible locations of \textit{Deep Impact}
liberated dust at these epochs.}
\label{kelley:fig:2}
\end{figure}

Synthetic images of the simulations, given a specific observer
geometry, predict the location of large particles that may have been
ejected as a result of the \textit{Deep Impact} collision.  We
examined the MIPS images from July and September~2005
(Fig.~\ref{kelley:fig:1}) to determine if slow moving, large dust
grains are lingering in the vicinity of the comet.
Figure~\ref{kelley:fig:2} presents sub-sections of the two MIPS images
and thermal emission contours (isothermal grains) from the large
particle simulations, assuming an ejected particle size distribution
of $dn/da \propto a^{-3.5}$.  The MIPS images have been unsharp masked
to enhance the coma asymmetries.  In the 2005 July MIPS image, there
is a coma feature extending in the anti-solar direction that is
coincident with the contours of the simulation.  These particles have
$\beta$ values ranging from $10^{-3}$ to $10^{-1}$ (lower $\beta$
values are found closer to the nucleus).  The 2005 September MIPS
image shows the same coma feature (in the anti-solar direction), but
the simulated \textit{Deep Impact} dust has rotated away from the
anti-solar direction.  We conclude that the anti-solar feature is a
young, normal coma feature and not due to \textit{Deep Impact}
liberated dust.  The limb brightened cone extending to the lower-right
from the nucleus is comprised of high velocity particles with $\beta
\lesssim 10^{-3}$ and does not appear to be present in the MIPS
images.  Possibly, the ambient coma is obscuring the \textit{Deep
Impact} ejected dust.  Observations and simulations of comet 9P in
September 2006 and January 2007 (when the coma emission has diminished
significantly) will be required to assess the existence of the largest
particles ($\beta = 10^{-4}-10^{-3}$).

\section{Conclusions}
The \textit{Deep Impact} collision with comet 9P/Tempel ejected small
particles from the comet nucleus, yet the upper size limit, and
therefore total ejected dust mass, is observationally not well
constrained.  Large slow moving grains ($a \gtrsim 100$~\mbox{$\mu$m})
in the ejecta plume (if they exist) have yet to be discovered.  We
simulate observations of large particles ejected by \textit{Deep
Impact} to predict the location of grains in \textit{Spitzer}/MIPS
images of comet 9P/Tempel at 30, 80, 420, and 560~days after the
\textit{Deep Impact} encounter.  The larger particles may be observed
as an enhancement to comet 9P/Tempel's trail in MIPS images 420 and
560~days after \textit{Deep Impact}.

This work is based on observations made with the \textit{Spitzer Space
Telescope}, which is operated by the Jet Propulsion Laboratory,
California Institute of Technology under a contract with NASA.
Support for this work was provided by NASA through an award issued by
JPL/Caltech.

%
%
%
%
%
%
%
%

%
%



\printindex
\end{document}